\documentclass{article}

\PassOptionsToPackage{numbers, compress}{natbib}

\usepackage[preprint]{neurips_2026} 

\usepackage[utf8]{inputenc} 
\usepackage[T1]{fontenc}    
\usepackage{hyperref}       
\usepackage{url}            
\usepackage{booktabs}       
\usepackage{amsfonts}       
\usepackage{nicefrac}       
\usepackage{microtype}      
\usepackage{xcolor}         
\usepackage{graphicx}       
\usepackage{float}          

\title{Passing: An Endless Journey through Reconstructed Spacetime with AI-Generated Sound}

\author{
  Akira Takahashi , Chihiro Nagashima , Zhi Zhong , Shusuke Takahashi , Yuki Mitsufuji\\
  Sony Group Corporation, Tokyo, Japan\\
}

\begin{document}

\maketitle

\begin{figure}[H]
    \centering
\includegraphics[width=0.75\linewidth]{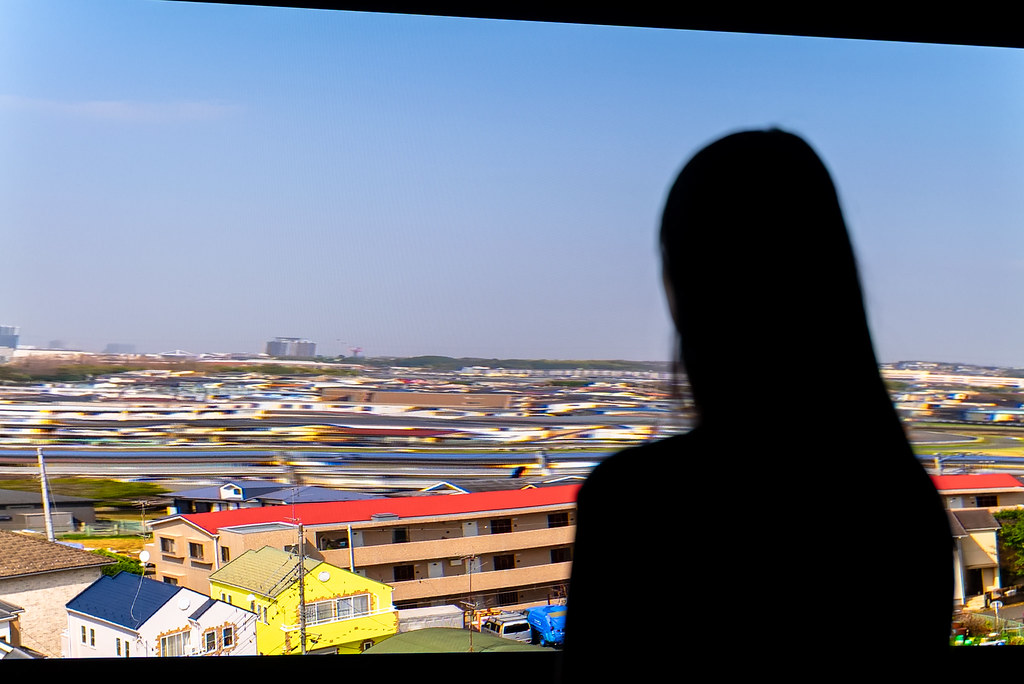}
    \caption{\emph{Passing}, a collaborative visual installation created with visual artist Ryu Furusawa, exhibited at NEORT++ in Tokyo from April 17–26, 2026 (\cite{furusawa2026passing}). 
}
    \label{fig:passing}
\end{figure}

\begin{abstract}
This paper introduces \textit{Passing}, an interactive audiovisual installation that generates an endless journey from a single continuous monorail-window recording by reconstructing it as a spatiotemporal volume. Rather than replaying the footage linearly, the work resamples its spatial and temporal structure along nonlinear trajectories, producing a continuously passing landscape whose depth, speed, and temporal order become unstable. A camera-based viewer-presence detection system estimates whether a viewer is present in the viewing zone and uses this presence state to influence transitions among rendered video sequences. The resulting video stream is fed into SpecMaskFoley, a real-time video-to-audio synthesis model that generates a synchronized soundscape for the reconfigured image. The model is not used to reconstruct an objectively correct soundtrack, but functions as a speculative listener, proposing a possible auditory interpretation of a world whose conventional spatial and temporal premises have been disrupted. \textit{Passing} distributes creative agency across the artist, who defines the rules of spacetime reconstruction; the AI model, which interprets the emergent visual flow as sound; and the audience, whose embodied presence influences the audiovisual trajectory. Through this structure, the work investigates how authorship and listening may be negotiated among human intention, machine inference, and audience interpretation.
Artwork page: \url{https://ryufurusawa.com/passing}
\end{abstract}

\section{Introduction}
\textit{Passing} is an audiovisual artwork that generates an endless journey from a single continuous monorail-window recording by digitally reconstructing its spacetime. This process transforms the familiar experience of ``passing’’ by deconstructing the perceptual foundations of linear time and three-dimensional space upon which it relies. This radical manipulation of the visual world raises a fundamental question: \emph{If the premises of homogeneous space and linear time have collapsed, how can sound exist, and how does listening come into being?}

Because the reconstructed video reorganizes the temporal order, motion, and spatial relations of the original recording, it no longer has a single physically correct acoustic counterpart. The task is therefore not to recover the sound of the original scene, but to generate an auditory interpretation that remains responsive to a visually coherent yet physically impossible flow.

To investigate this task, we employed SpecMaskFoley~\cite{zhong2025specmaskfoley}, a video-to-audio (V2A) synthesis model, not to reproduce an objectively correct soundtrack, but to hear how an AI model with no body of its own may respond to reconfigured spacetime. The AI acts as a speculative listener, generating sound anew for each moment of the nonlinear video and proposing an auditory experience for a world unbound by conventional physics.
Creative agency is thereby redistributed among the artist, who designs the rules of spacetime reconstruction; the AI, which interprets the emergent visual flow as sound; and the audience, whose presence or absence in the viewing zone influences the visual path.
\section{Related Works}

\subsection{Spatiotemporal Manipulation in Media Art}

The visual methodology of \textit{Passing} belongs to a lineage of media-art practices that reorganize recorded images by altering the relation between space and time. Digital slit-scan practices, as catalogued by Golan Levin~\cite{levin_slitscan_catalogue}, and broader discussions of space-time correlations in moving-image media by Susanne Jaschko~\cite{jaschko_spacetime_2003}, provide a historical and theoretical context for treating video not as a fixed sequence of frames, but as a structure in which spatial and temporal dimensions can be exchanged, stretched, or reconfigured. Interactive works such as Alvaro Cassinelli's \textit{Khronos Projector}~\cite{cassinelli_khronos_2004} and Camille Utterback's \textit{Liquid Time}~\cite{utterback_liquid_time_2000} further demonstrate how the viewer's present action can be incorporated into the temporal structure of recorded video.

A more direct artistic precedent within Ryu Furusawa's practice is \textit{Mid Tide \#3}~\cite{furusawa2024midtide3}, which uses high-resolution, high-frame-rate footage of waves as source material to make altered spacetime perceptible. \textit{Passing} extends this lineage by treating a continuous monorail-window recording as a spatiotemporal volume and generating new moving images along nonlinear cross-sectional trajectories. Unlike earlier interactive temporal manipulations constrained by real-time high-resolution rendering, \textit{Passing} prepares transformed video clips in advance and reorganizes their playback in response to viewer presence. By coupling this presence-mediated visual sequence with real-time AI sound, \textit{Passing} connects spatiotemporal image reconstruction, machine listening, and distributed creative agency.

\subsection{Speculative Video-to-Audio Synthesis}

The auditory component of \textit{Passing} draws on recent advances in V2A synthesis.
Several recent V2A methods have made significant progress in generating plausible sounds from visual or multimodal input~\cite{cheng2025taming,jeong2024rewas,xing2024seeingandhearing,viertola2025vaura,wang2024v2amapper,zhang2024foleycrafter,liu2024vatt,wang2024frieren}. Among them, MMAudio~\cite{cheng2025taming} demonstrates strong audiovisual synchronization and high-quality sound generation from video. SpecMaskFoley~\cite{zhong2025specmaskfoley} also achieves synchronized V2A generation while supporting real-time inference, making it suitable for interactive installation settings.

In contrast, \textit{Passing} uses V2A synthesis as a speculative interpretive process. The aim is not to reconstruct the correct sound of a monorail scene, but to ask what kind of sound an AI model may infer from a visual world whose temporal and spatial premises have been reconfigured. In this context, synchronization supports artistic interpretation rather than objective acoustic reconstruction. Instead of being conditioned on the original footage in its recorded order, the model receives a spatiotemporally reconfigured video stream and generates a possible auditory interpretation of the altered visual world.

\subsection{Distributed Agency in AI Art}

The conceptual framework of \textit{Passing} is related to discussions of distributed agency and authorship in AI-based creative practice. As AI systems enter artistic production, authorship is increasingly understood not as the property of a single creator, but as something negotiated among human and nonhuman actors~\cite{audry2021aiart}. 
Our previous project, the generative sound installation \emph{Studies for}~\cite{evala2024studiesfor, chihiro2025studiesfor}, examined this negotiation through an AI archive trained on sound artist evala's past works. 
Other works, such as Suzanne Kite's \emph{Listener}~\cite{kite2018listener} and Justine Emard's \emph{Co(AI)xistence}~\cite{emard2017coaixistence}, have explored AI or artificial agents as collaborators, others, or interlocutors.

\textit{Passing} extends this discussion by explicitly including the audience in the agency structure. The artist defines the rules of spacetime reconstruction, the AI model interprets the resulting visual flow as sound, and the audience indirectly influences the visual path through their presence or absence in the viewing zone. The work therefore constructs a three-way agency loop in which audience behavior shapes the material that the AI interprets in real time.

\section{Artwork and Interaction Design}

\subsection{Reconstructing Spacetime from a Continuous Recording}
\label{ssec:3_1}

\begin{figure}[t]
  \centering
  \includegraphics[width=0.9\linewidth]{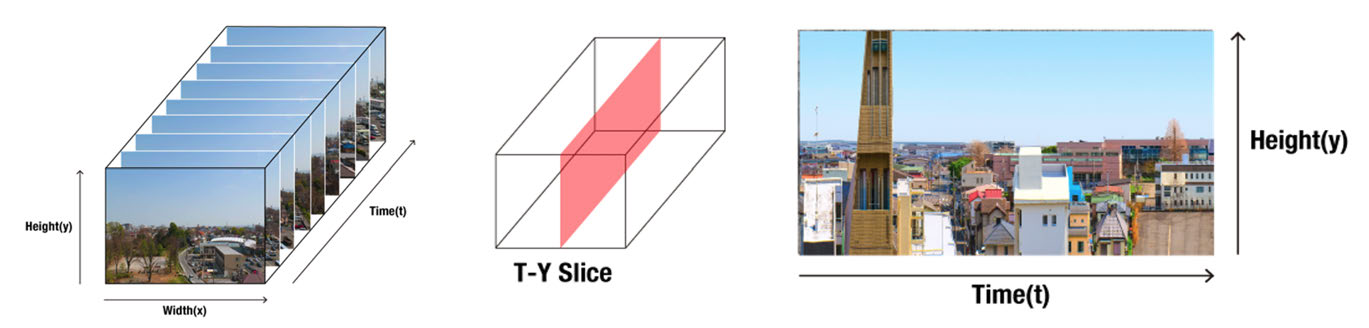}
  \caption{
  Spacetime reconstruction in \textit{Passing}. 
Frames are stacked along the temporal axis to form a spatiotemporal volume. 
Ordinary playback follows this axis, whereas \textit{Passing} generates new moving images by shifting and rotating the cross-sectional plane. 
As an illustrative example, a 90$^\circ$ rotation exchanges the temporal axis with the horizontal image dimension.
  }
  \label{fig:spacetime_reconstruction}
\end{figure}

\textit{Passing} is built from a single continuous 4K 480-fps HDR recording captured on March 27, 2025, through the window of the Tama Monorail between Tachikawa and Tama Center using a high-resolution, high-speed camera. Rather than treating this recording as a fixed sequence of frames to be played back linearly, the work treats the moving image as a spatiotemporal volume. Each frame is regarded as a photographic slice, and the sequence of frames is stacked along the temporal axis to form a virtual volume that can be traversed in ways that could not have existed in the captured scene itself. Once this volume is constructed, cross-sections can in principle be extracted at arbitrary angles. By continuously shifting and rotating the cross-sectional plane along selected trajectories, the system produces moving images that are no longer bound to the original frame order. In this sense, the transformation does not simply edit the recorded video; it resamples the temporal and spatial structure already contained within it. Fig.~\ref{fig:spacetime_reconstruction} schematically illustrates this principle using a 90$^\circ$ cross-sectional rotation as an example. 

The recorded landscape contains multiple latent movements, including the forward movement of the train, the relative motion of nearby and distant objects, and the local motion of people, cars, and other elements within the scene. Depending on the angle of the extracted cross-section, some movements may appear to be suspended, whereas others may be stretched, reversed, or displaced. Therefore, the transformation exposes multiple temporalities latent within the original footage.

For the installation, multiple trajectory variants were generated by varying parameters such as departure timing and rotational speed, then pre-rendered and connected into a continuous playback structure. Each trajectory preserves the right-to-left flow of the passing landscape, including combinations of horizontal inversion and reverse playback. In this way, \textit{Passing} produces an endless journey from a finite source recording by reorganizing, rather than synthetically generating, the landscape. 

\subsection{Audience-Mediated Sequence Reordering}

In \textit{Passing}, viewer interaction does not directly control the sound. The viewer-presence estimate is used only for video sequence control; the resulting visual sequence is then passed to the real-time sound generation model. Thus, interaction is mediated through the image: the viewer affects the visual path, and the AI model generates sound in response to the altered visual flow. 

The system detects whether a viewer is present in the viewing zone and uses this state to probabilistically enable transition branches among transformed video sequences. Each sequence follows a different cross-sectional trajectory through the same spatiotemporal volume, so the work does not simply play a predetermined loop but shifts its temporal path in response to embodied presence. 
The rendered video content is organized into primary sequence groups and transition branches, allowing the system to sustain an endless video loop while viewer presence probabilistically enables branch transitions. The detailed group structure and sequence-transition graph are provided in Appendix~\ref{app:sequence_design} and Fig.~\ref{fig:sequence_transition_appendix}.

Changes in the displayed video sequence alter the temporal and spatial features available to the sound synthesis model, including motion direction, speed, event density, transition rhythm, and contradictions produced by inversion or reverse playback. Thus, the viewer affects sound only through the image, altering the world that the AI model listens to rather than operating sound as an instrument. 

\subsection{Installation Loop}

\begin{figure}[t]
  \centering
  \includegraphics[width=0.8\linewidth]{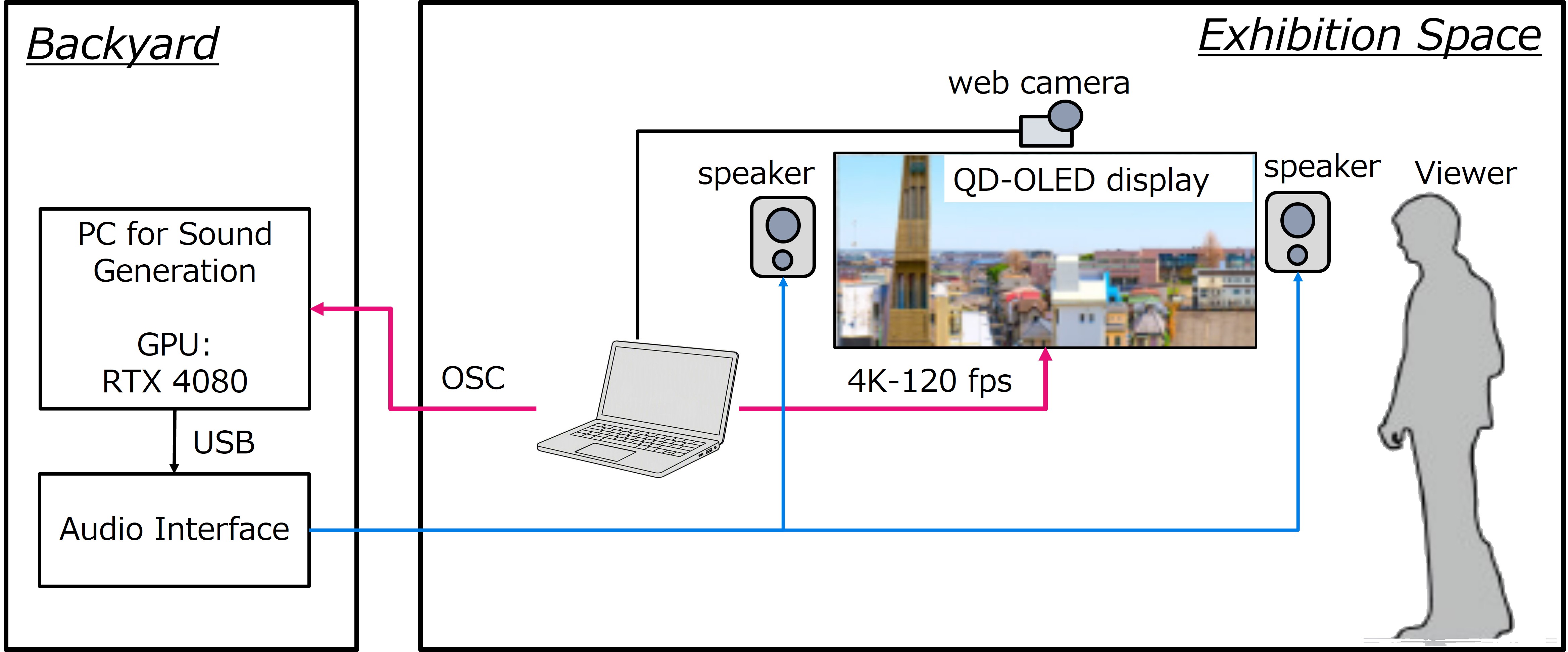}
  \caption{Installation view and system configuration of \textit{Passing}. 
The work is presented on a 4K 120-fps display with stereo sound reproduction. 
A camera estimates viewer presence in front of the display and sends the presence state to the video playback control system, which selects rendered video clips. 
The selected clip determines the corresponding visual-conditioning sequence provided to SpecMaskFoley running on an RTX 4080 GPU, and the generated stereo audio is delivered through an audio interface and loudspeakers.}
  \label{fig:equipment}
\end{figure}

The installation is implemented as an interactive loop connecting viewer-presence sensing, video sequence control, and real-time sound synthesis, as illustrated in Fig.~\ref{fig:equipment}. A camera-based viewer-presence detection system estimates whether a viewer is present in front of the display and sends only this presence state to the playback control system; no camera imagery is recorded or stored.

When a transition is triggered, the control system sends an OSC (Open Sound Control) cue to the playback engine, which selects the next clip from the rendered dataset. The selected video is displayed as 4K HDR content at 120 fps, while the corresponding visual-conditioning inputs are provided to SpecMaskFoley running on an RTX 4080 workstation. SpecMaskFoley synthesizes a synchronized stereo soundscape for the selected visual path, which is routed through an audio interface and loudspeakers. In this configuration, the viewer does not directly operate the sound-generation model. Instead, viewer presence changes the visual sequence, and the AI model generates sound in response to the resulting visual flow.

\section{Real-time Sound Generation}

The auditory component of \textit{Passing} is generated in real time by SpecMaskFoley~\cite{zhong2025specmaskfoley}, a V2A synthesis model capable of producing synchronized sound for the nonlinear visual flow. 
Building on our previous project on continuous real-time sound generation under text/audio conditioning~\cite{chihiro2025studiesfor}, \textit{Passing} extends this paradigm to video-conditioned generation, where the conditioning signal is a dynamically reordered visual stream produced by reconstructed spacetime.
This section summarizes the model configuration, training data, and deployment setup used in the installation.

\subsection{Model Architecture and Control}
SpecMaskFoley uses a pre-trained SpecMaskGIT~\cite{comunita2024specmaskgit} model as the audio backbone, together with a ControlNet branch and an FT-Aligner that adapts SynchFormer-based~\cite{synchformer2024iashin} temporal video features to the time-frequency ControlNet implementation. The ControlNet branch provides temporal alignment between the visual and auditory streams, allowing the model to generate sounds synchronized with both sustained visual motion and instantaneous events. A schematic overview of the architecture is provided in Appendix~\ref{app:specmaskfoley_architecture}.

For \textit{Passing}, we omitted the CLIP visual encoder used in the original SpecMaskFoley configuration. The final system therefore relies on ControlNet for temporal video alignment and on CLAP-based audio conditioning for semantic guidance. Ambient sounds recorded inside the Tama Monorail were encoded with a pre-trained CLAP~\cite{laion2023clap} audio encoder and used as conditioning information for the SpecMaskGIT backbone. This configuration supports the artistic aim of letting the audio domain guide the model's interpretation of reconstructed spacetime, while simplifying the conditioning pathway for real-time deployment.

\subsection{Training Data and Adaptation}
The training data for SpecVQGAN~\cite{iashin2021SpecVQGAN}, SpecMaskGIT, and Vocos~\cite{siuzdak2023vocos} consist of the large-scale audio event datasets AudioSet~\cite{gemmeke2017audioset} and VGGSound~\cite{chen2020vggsound}. As a preprocessing step, we applied source separation to remove human speech components to reduce the risk of unintentionally synthesizing human-like voices.

SpecMaskFoley was developed by fine-tuning the pre-trained SpecMaskGIT model. In addition to VGGSound, the training dataset was augmented with approximately 35 min of custom audiovisual data recorded during the Tama Monorail shoot. To compensate for the limited size of this custom dataset, we applied spatial crops to the video, channel manipulations to the audio, and temporally reversed audiovisual pairs while preserving synchronization.

\subsection{Real-time Deployment}
\label{ssec:4_3}
Real-time generation is essential for \textit{Passing}, because the sound must be synthesized anew for each moment of the nonlinear video loop rather than being replayed from prepared audio. For deployment, the displayed video and the visual-conditioning input were separated. The rendered clips were displayed as 4K HDR video at 120 fps, while corresponding conditioning images were precomputed and stored at 224 $\times$ 224 pixels and 25 fps. This deployment strategy allowed SpecMaskFoley to synthesize audio continuously for the selected visual path without extracting visual features from each displayed 4K 120-fps frame at runtime.

SpecMaskFoley was adopted because it supported continuous synchronized audio generation on the installation workstation while producing output suitable for the exhibition environment. Before the exhibition, on-site stability tests and listening checks with the artist confirmed that the A/B loop retained a stable train-interior atmosphere while the C/D transition sequences produced stronger sonic variation in response to the more distorted visual flow, guiding the final balance among synchronization, fidelity, and naturalness. In the final setup, the system generated 32-kHz stereo audio in real time and routed it through the exhibition sound system.
\section{Discussion}

\subsection{Mediated Observation and Listening}

\textit{Passing} redistributes creative agency through mediation rather than direct control. The artist defines the recording, transformation rules, and constraints of the system; the viewer's presence influences the visual path; and the AI model generates sound in response to the resulting nonlinear visual flow. 
Crucially, the viewer does not operate the sound as an instrument: the presence state is not mapped directly onto volume, pitch, or timbre, but instead alters the visual world that the model receives.

This mediation destabilizes both observation and listening. Because the moving image is generated by traversing a spatiotemporal volume rather than replaying a fixed sequence of frames, the landscape remains recognizable while depth, speed, viewpoint, and temporal order no longer coincide consistently. The audio is likewise not a prerecorded track attached to a stable viewpoint, but is synthesized in real time from the currently displayed visual flow. In the stable A/B sequence loop, the generated sound tended to maintain a continuous train-interior atmosphere, whereas the C/D transition sequences allowed stronger sonic variation in response to the more distorted visual flow. The artist described this effect as a sensation that the ``point of observation wavers, as if in an out-of-body state.'' We interpret this as a loosening of the usual correspondence between the visual point of view, the listening position, the screen, and the viewer's body. In this sense, the AI model does not simply add sound to an image; it provides an auditory interpretation for a world in which the conditions of observation have already been destabilized.

\subsection{Negotiated Soundscape and Relational Authorship}

The final soundscape was developed through iterative collaboration with the artist. The aim was not to produce a realistic soundtrack or a direct sonification of visual parameters, but to hear how the AI might interpret motion, speed, and temporal instability in a world where time and space have collapsed. Adjustments to generation parameters and to the balance among fidelity, synchronization, and naturalness defined the conditions under which the model's inference could become artistically meaningful.

The model's limitations are central to this negotiation. Because it responds to visual motion and temporal change without possessing a body or lived spatial experience, its output may contain mismatches, ambiguities, or unexpected correspondences between image and sound. Rather than treating these as failures, \textit{Passing} treats them as productive frictions through which the model's interpretive agency becomes perceptible. Creative agency therefore emerges not as a property of a single actor, but as mediated, conditional, and relational.

\section{Conclusion}
This paper introduced \textit{Passing}, an audiovisual artwork that generates an endless journey from a single continuous recording by digitally reconstructing its spacetime. By employing SpecMaskFoley, a real-time V2A synthesis model, we explored a fundamental question: \emph{If the premises of homogeneous space and linear time have collapsed, how can sound exist, and how does listening come into being?} The system was designed not to reproduce a correct answer, but to hear the response of an AI model with no body of its own, positioning the model as an active participant in the creative process.

Our primary contribution lies in the investigation of distributed creative agency. We have presented a framework in which authorship is negotiated among three distinct agents: the artist, who designs the fundamental rules of the world; the AI, which proposes a real-time auditory interpretation from within that world; and the audience, whose presence influences the emergent audiovisual path and ultimately constructs personal meaning. This tripartite model moves beyond a simple human-tool relationship, revealing a dynamic and constantly negotiated process of shared imagination. \textit{Passing} serves as a practical case study of how creative agency can be productively redistributed, rather than merely automated or diminished, through collaboration with AI.

Rather than treating real-time V2A synthesis as an isolated module, \textit{Passing} integrates it into a continuous exhibition-scale audiovisual system in which visual reconstruction, machine inference, and audience presence mutually condition one another during live playback.

\clearpage
\section*{Acknowledgements}
We are deeply grateful to visual artist Ryu Furusawa for giving us the opportunity to collaborate on his visual work \textit{Passing} and to apply our sound-generation technology within his artwork. 
In particular, the challenge of generating sound in real time for an endless audiovisual journey through reconfigured spacetime has become an important direction that we hope to continue exploring in future work.

We also sincerely thank Hiromasa Hosotani of Sony Corporation for facilitating our introduction to visual artist Ryu Furusawa. 
For our research team, this collaboration began in February 2026, and Mr. Hosotani's support and contributions leading up to the exhibition in mid-April were highly significant. We deeply appreciate these contributions.

\bibliographystyle{unsrtnat}
\bibliography{refs26}

@misc{furusawa2026passing,
  author = {Ryu Furusawa},
  title = {Passing},
  howpublished = {\url{https://ryufurusawa.com/passing}},
  note = {Accessed: 2026-07-28},
  year = {2026},
  institution = {NEORT++, Tokyo}
}

@misc{furusawa2024midtide3,
  author = {Ryu Furusawa},
  title = {{Mid Tide \#3}},
  howpublished = {\url{https://ryufurusawa.com/midtide3}},
  note = {Accessed: 2026-07-28},
  year = {2024},
  institution = {NTT InterCommunication Center (ICC), Tokyo}
}

@misc{evala2024studiesfor,
  author = {Evala},
  title = {Studies for},
  howpublished = {\url{https://www.ntticc.or.jp/en/archive/works/studies-for/}},
  note = {Accessed: 2026-07-28},
  year = {2024},
  institution = {NTT InterCommunication Center (ICC), Tokyo}
}

@inproceedings{chihiro2025studiesfor,
  title={{\textquoteleft}{Studies for}{\textquoteright}: A Human-{AI} Co-Creative Sound Artwork Using a Real-time Multi-channel Sound Generation Model},
  author={Nagashima, Chihiro and Takahashi, Akira and Zhong, Zhi and Takahashi, Shusuke and Mitsufuji, Yuki},
  booktitle={NeurIPS Creative AI Track 2025},
  year={2025}
}

@book{audry2021aiart,
  author = {Sofian Audry},
  title = {Art in the Age of Machine Learning},
  year = {2021},
  publisher = {MIT Press}
}

@misc{kite2018listener,
  author = {Kite, Suzanne},
  title = {Listener},
  year = {2018},
  howpublished = {\url{https://www.kitekitekitekite.com/portfolio/listener}},
  note = {Accessed: 2026-07-28}
}

@misc{emard2017coaixistence,
  author = {Emard, Justine},
  title = {{Co(AI)xistence}},
  year = {2017},
  howpublished = {\url{https://justineemard.com/coaixistence-2/}},
  note = {Accessed: 2026-07-28}
}

@misc{levin_slitscan_catalogue,
  author       = {Levin, Golan},
  title        = {An Informal Catalogue of Slit-Scan Video Artworks and Research},
  howpublished = {\url{http://www.flong.com/archive/texts/lists/slit_scan/index.html}},
  note         = {Accessed: 2026-07-28},
  year         = {2003}
}

@incollection{jaschko_spacetime_2003,
  author    = {Jaschko, Susanne},
  title     = {Space-Time Correlations Focused in Film Objects and Interactive Video},
  booktitle = {Future Cinema: The Cinematic Imaginary after Film},
  publisher = {MIT Press},
  address   = {Cambridge, MA},
  year      = {2003}
}

@misc{cassinelli_khronos_2004,
  author       = {Cassinelli, Alvaro},
  title        = {{Khronos Projector}},
  year         = {2004},
  howpublished = {\url{https://alvarocassinelli.com/khronos-projector/}},
  note         = {Accessed: 2026-07-28}
}

@misc{utterback_liquid_time_2000,
  author       = {Utterback, Camille},
  title        = {{Liquid Time}},
  year         = {2000},
  howpublished = {\url{https://camilleutterback.com/projects/liquid-time-series/}},
  note         = {Accessed: 2026-07-28}
}

@inproceedings{zhong2025specmaskfoley,
  title={{SpecMaskFoley}: Steering Pretrained Spectral Masked Generative Transformer Toward Synchronized Video-to-audio Synthesis via {ControlNet}},
  author={Zhong, Zhi and Takahashi, Akira and Cui, Shuyang and Toyama, Keisuke and Takahashi, Shusuke and Mitsufuji, Yuki},
  booktitle={WASPAA},
  year={2025}
}

@inproceedings{liu2024vatt,
  title={Tell What You Hear From What You See--Video to Audio Generation Through Text},
  author={Liu, Xiulong and Su, Kun and Shlizerman, Eli},
  booktitle={NeurIPS},
  year={2024}
}

@inproceedings{wang2024frieren,
  title={Frieren: Efficient video-to-audio generation network with rectified flow matching},
  author={Wang, Yongqi and Guo, Wenxiang and Huang, Rongjie and Huang, Jiawei and Wang, Zehan and You, Fuming and Li, Ruiqi and Zhao, Zhou},
  booktitle={NeurIPS},
  year={2024}
}

@inproceedings{viertola2025vaura,
  title={Temporally aligned audio for video with autoregression},
  author={Viertola, Ilpo and Iashin, Vladimir and Rahtu, Esa},
  booktitle={ICASSP},
  year={2025},
}

@article{zhang2024foleycrafter,
  title={Foleycrafter: Bring silent videos to life with lifelike and synchronized sounds},
  author={Zhang, Yiming and Gu, Yicheng and Zeng, Yanhong and Xing, Zhening and Wang, Yuancheng and Wu, Zhizheng and Chen, Kai},
  journal={arXiv preprint arXiv:2407.01494},
  year={2024}
}

@inproceedings{wang2024v2amapper,
  title={V2a-mapper: A lightweight solution for vision-to-audio generation by connecting foundation models},
  author={Wang, Heng and Ma, Jianbo and Pascual, Santiago and Cartwright, Richard and Cai, Weidong},
  booktitle={AAAI},
  year={2024}
}

@inproceedings{cheng2025taming,
  title={{MMAudio}: Taming Multimodal Joint Training for High-Quality Video-to-Audio Synthesis},
  author={Cheng, Ho Kei and Ishii, Masato and Hayakawa, Akio and Shibuya, Takashi and Schwing, Alexander and Mitsufuji, Yuki},
  booktitle={CVPR},
  year={2025}
}

@article{jeong2024rewas,
  title={Read, watch and scream! sound generation from text and video},
  author={Jeong, Yujin and Kim, Yunji and Chun, Sanghyuk and Lee, Jiyoung},
  journal={arXiv preprint arXiv:2407.05551},
  year={2024}
}

@inproceedings{xing2024seeingandhearing,
  title={Seeing and hearing: Open-domain visual-audio generation with diffusion latent aligners},
  author={Xing, Yazhou and He, Yingqing and Tian, Zeyue and Wang, Xintao and Chen, Qifeng},
  booktitle={CVPR},
  year={2024}
}

@inproceedings{synchformer2024iashin,
  title     = {{SynchFormer}: Efficient Synchronization from Sparse Cues},
  author    = {Iashin, Vladimir and Xie, Weidi and Rahtu, Esa and Zisserman, Andrew},
  booktitle = {ICASSP},
  year      = {2024}
}

@inproceedings{comunita2024specmaskgit,
  title={{SpecMaskGIT}: Masked generative modeling of audio spectrograms for efficient audio synthesis and beyond},
  author={Comunit{\`a}, Marco and Zhong, Zhi and Takahashi, Akira and Yang, Shiqi and Zhao, Mengjie and Saito, Koichi and Ikemiya, Yukara and Shibuya, Takashi and Takahashi, Shusuke and Mitsufuji, Yuki},
  booktitle={ISMIR},
  year={2024}
}

@article{siuzdak2023vocos,
  title={Vocos: Closing the gap between time-domain and {Fourier-based} neural vocoders for high-quality audio synthesis},
  author={Siuzdak, Hubert},
  journal={arXiv preprint arXiv:2306.00814},
  year={2023}
}

@article{iashin2021SpecVQGAN,
  title={Taming visually guided sound generation},
  author={Iashin, Vladimir and Rahtu, Esa},
  journal={arXiv preprint arXiv:2110.08791},
  year={2021}
}

@inproceedings{laion2023clap,
  author={Wu, Yusong and Chen, Ke and Zhang, Tianyu and Hui, Yuchen and Berg-Kirkpatrick, Taylor and Dubnov, Shlomo},
  booktitle={ICASSP}, 
  title={Large-Scale Contrastive Language-Audio Pretraining with Feature Fusion and Keyword-to-Caption Augmentation}, 
  year={2023},
}

@inproceedings{gemmeke2017audioset,
  title={{Audio Set}: An ontology and human-labeled dataset for audio events},
  author={Gemmeke, Jort F and Ellis, Daniel PW and Freedman, Dylan and Jansen, Aren and Lawrence, Wade and Moore, R Channing and Plakal, Manoj and Ritter, Marvin},
  booktitle={ICASSP},
  year={2017},
}

@inproceedings{chen2020vggsound,
  title={{VGGSound}: A large-scale audio-visual dataset},
  author={Chen, Honglie and Xie, Weidi and Vedaldi, Andrea and Zisserman, Andrew},
  booktitle={ICASSP},
  year={2020}
}


\newpage

\appendix
\renewcommand{\thesection}{Appendix \Alph{section}} \renewcommand{\thesubsection}{\Alph{section}.\arabic{subsection}}
\section{Technical Details of Real-time Video-to-Audio Generation}
\subsection{SpecMaskFoley Configuration and Architectural Modifications}
\label{app:specmaskfoley_architecture}

\begin{figure}[H]
  \centering
  \includegraphics[width=0.9\linewidth]{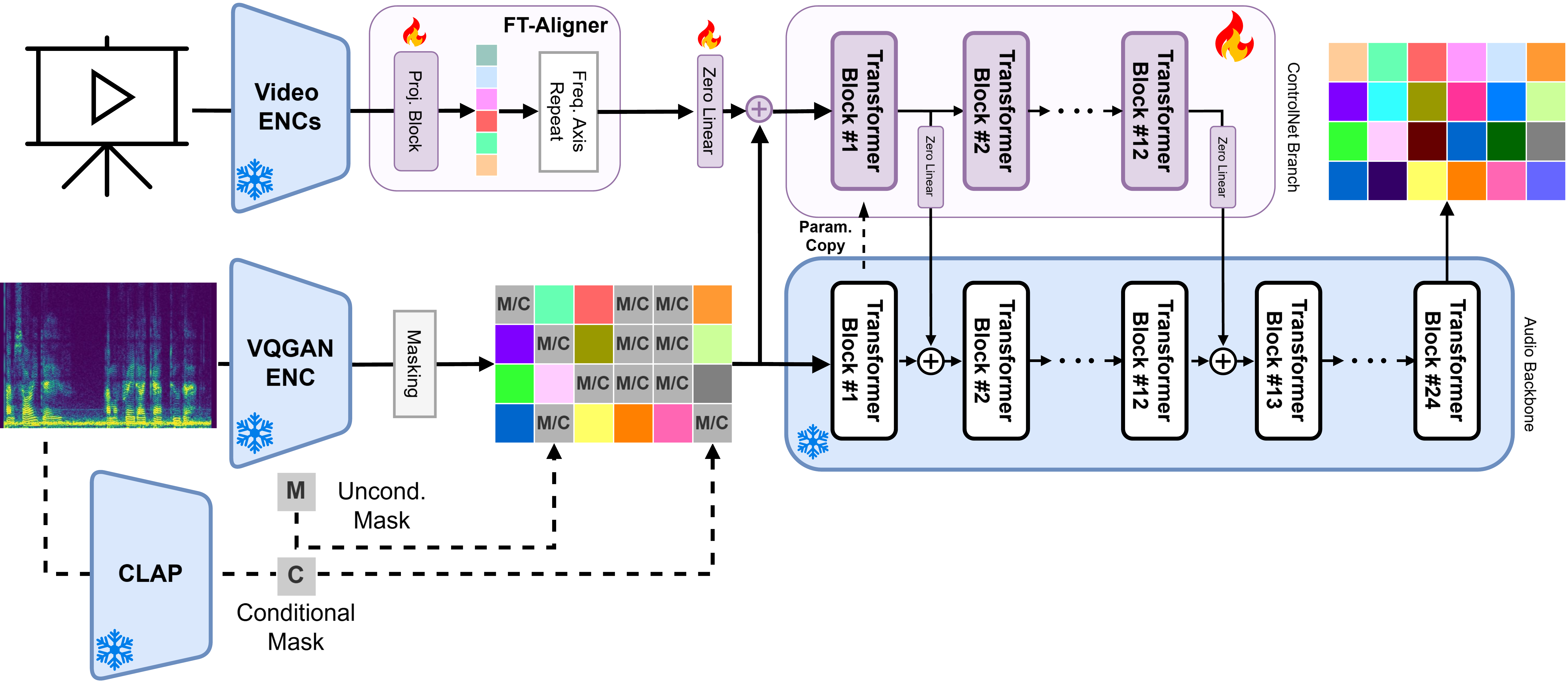}
  \caption{Overview of SpecMaskFoley. Ice icons indicate frozen modules, and fire icons indicate trainable modules. A CLAP embedding is treated as a conditional mask following SpecMaskGIT. In \textit{Passing}, ControlNet provides temporal alignment, while CLAP-based conditioning provides semantic guidance.}
  \label{fig:specmaskfoley}
\end{figure}
This section provides additional details on the modifications made to the original SpecMaskFoley architecture~\cite{zhong2025specmaskfoley} for its application in \textit{Passing}.
Fig.~\ref{fig:specmaskfoley} shows the architecture, in which ControlNet provides temporal video alignment and CLAP-based conditioning provides semantic guidance.

While the original SpecMaskFoley model utilizes both a ControlNet for temporal synchronization and a CLIP visual encoder for semantic understanding of the video content, we deliberately omitted the CLIP visual encoder in this work for two primary reasons.

First, from an artistic standpoint, the decision was made to entrust semantic conditioning entirely to the auditory domain. Using only the CLAP audio encoder with ambient sounds recorded inside the Tama Monorail as conditioning, we anchored the generated soundscape to the stable acoustic context of a train interior. This conditioning was not intended to reconstruct the original monorail soundtrack, but to prevent the output from drifting toward unrelated or excessively artificial textures while still allowing the model to respond to the reconfigured visual motion. In this configuration, the model interprets the visual flow through a sonic rather than explicitly visual semantic context.

Second, from a technical perspective, the CLIP visual encoder proved to be a significant bottleneck for real-time inference. Its removal was crucial to ensure the stable, low-latency performance required for the interactive installation, where sound must be synthesized anew for the evolving video stream without perceptible delay. Therefore, the final architecture relies on ControlNet for temporal alignment and CLAP for semantic guidance, a configuration that satisfies both the artistic goals and technical requirements of this project.

For deployment, the video clips were converted in advance into visual-conditioning image sequences at 224 $\times$ 224 pixels and 25 fps. These images were cached before playback and loaded according to the selected sequence, avoiding the need to process every displayed 4K 120-fps frame during the installation.

In the \textit{Passing} configuration, the 32 kHz SpecVQGAN tokenizer, SpecMaskGIT audio backbone, and Vocos vocoder were trained on AudioSet and VGGSound before the SpecMaskFoley adaptation, which additionally used the custom Tama Monorail recordings.

\begin{table}[H]
  \caption{A comparison of the model details}
  \label{optimization}
  \centering
  \begin{tabular}{c|cc}
    \toprule
         & SpecMaskFoley (\cite{zhong2025specmaskfoley})     & Passing \\
    \midrule
    V2A Adaptation Dataset & VGGSound 500 hours & VGGSound 500 hours \\
    & & + Tama Monorail recording data \\
    Sampling rate & 22.05 kHz & 32 kHz\\
    SpecVQGAN & params: 72M & params: 72M  \\
     & 80 Mel bins $\times$ 848 frames & 128 Mel bins $\times$ 800 frames\\ 
     & 265 tokens (5 $\times$ 53) & 400 tokens (8 $\times$ 50) \\
     & 820$\times$ compression & 800$\times$ compression \\
    SpecMaskFoley & params: 300M  & params: 300M \\
     Main Network & 24 Transformer blocks & 24 Transformer blocks \\
     & 768 dim 8 heads attention & 768 dim 8 heads attention \\
     ControlNet  & 12 Transformer blocks & 12 Transformer blocks \\
     Vocoder & HiFi-GAN & Vocos \\
    \bottomrule
  \end{tabular}
\end{table}

\subsection{Comparison of Video-to-Audio Synthesis Performance}
\label{app:v2a_performance}

\begin{figure}[H]
    \centering
    \includegraphics[width=0.8\linewidth]{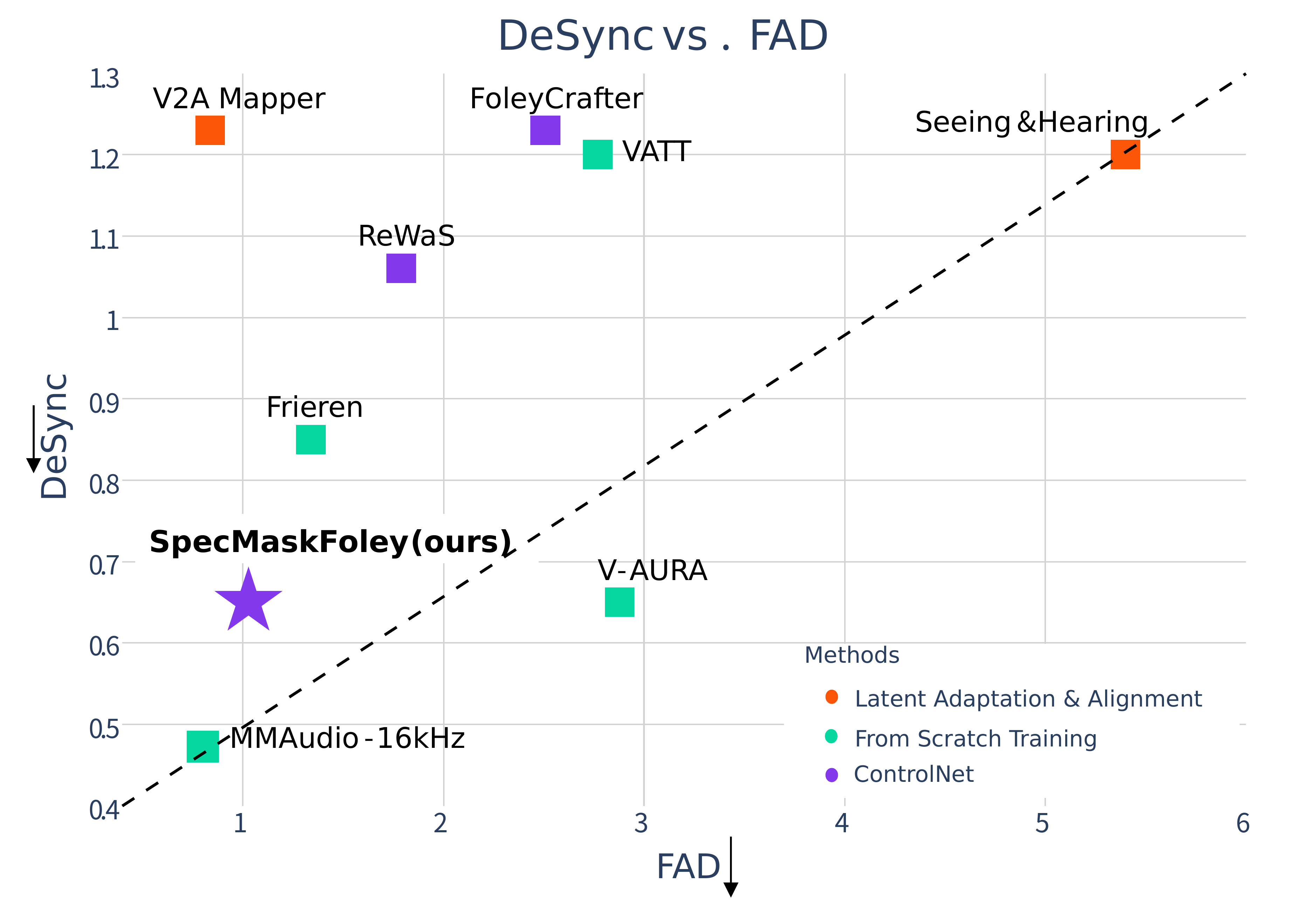}
    \caption{Audio synthesis quality and audio-video synchronization performance of representative V2A methods on the VGGSound test dataset, adapted from SpecMaskFoley~\cite{zhong2025specmaskfoley}. FAD measures audio quality and DeSync measures temporal alignment; lower is better for both. SpecMaskFoley achieves competitive performance without from-scratch V2A training, motivating its use in \textit{Passing}.}
    \label{fig:v2a_performance}
\end{figure}

Fig.~\ref{fig:v2a_performance} summarizes the reported performance of SpecMaskFoley in comparison with representative V2A synthesis methods on the VGGSound test dataset. The comparison follows the evaluation in SpecMaskFoley~\cite{zhong2025specmaskfoley}, where audio synthesis quality is measured by Fréchet Audio Distance (FAD) and audiovisual temporal alignment is measured by DeSync score. Lower values are better for both metrics.

The figure is included to contextualize the choice of SpecMaskFoley as the real-time sound-generation model used in \textit{Passing}. Unlike from-scratch V2A models, SpecMaskFoley adapts a pre-trained SpecMaskGIT audio generation backbone through a ControlNet branch and a frequency-aware temporal feature aligner. This design allows the model to achieve competitive audio quality and temporal synchronization while supporting efficient inference, making it suitable for interactive installation settings. The comparison is not intended as an evaluation of the final soundscape of \textit{Passing}, but as background evidence for the model's suitability as a real-time V2A component.

\section{Video Reconstruction and Sequence Design}
\label{app:video_reconstruction}

\subsection{Digital Coordinate Transformation}
\label{app:coordinate_transformation}

As described in Section~\ref{ssec:3_1} and illustrated in Fig.~\ref{fig:spacetime_reconstruction}, \textit{Passing} treats the recorded moving image as a spatiotemporal volume by stacking frames along the temporal axis. Ordinary playback corresponds to a cross-sectional plane advancing through this volume in the original frame order. In contrast, \textit{Passing} changes the position and angle of this plane over time, extracting images in which spatial and temporal dimensions are partially exchanged, stretched, or reversed.

This coordinate transformation changes the perceptual structure of the image without synthesizing a new landscape. As the cross-section tilts away from the ordinary frame orientation, the simultaneity normally contained within a single frame begins to collapse, and motion traces embedded in the recording become visible. Consequently, depth, speed, viewpoint, and temporal order no longer remain consistently aligned, even though the image retains the visual continuity of recorded scenery.

This section defines the coordinate model underlying the reconstruction. The following section describes how this model was instantiated as a set of cross-sectional trajectories for the installation.

\subsection{Cross-sectional Trajectory Generation}
\label{app:trajectory_generation}

Based on the coordinate transformation described above, we generated multiple cross-sectional trajectories for the installation. Each trajectory defines how the virtual slicing plane moves through the spatiotemporal volume over time, rather than how the original frames are replayed in their recorded order.

The trajectory variants were created by varying parameters such as the departure timing and rotational speed of the cross-sectional plane. These parameters determine how different latent movements in the recorded scene become visible, including the forward movement of the train, the relative motion of near and distant objects, and the local motion of people, cars, or other elements. Depending on the trajectory, these movements may appear suspended, stretched, reversed, or displaced.

A key constraint in generating these trajectories is the preservation of the overall right-to-left flow of the landscape, including cases in which horizontal inversion is combined with reverse playback. This constraint ensures that the visual experience remains connected to the bodily perception of passing scenery, even when the internal temporal structure of the image no longer follows linear time.

The resulting trajectories were pre-rendered as video clips for use in the installation playback system. Therefore, the work does not synthesize a new landscape from scratch; rather, it reorganizes temporal and spatial structures already present in the recorded footage.

\subsection{Sequence Groups and Transition Graph}
\label{app:sequence_design}

\begin{figure}
  \centering
  \includegraphics[width=1.0\linewidth]{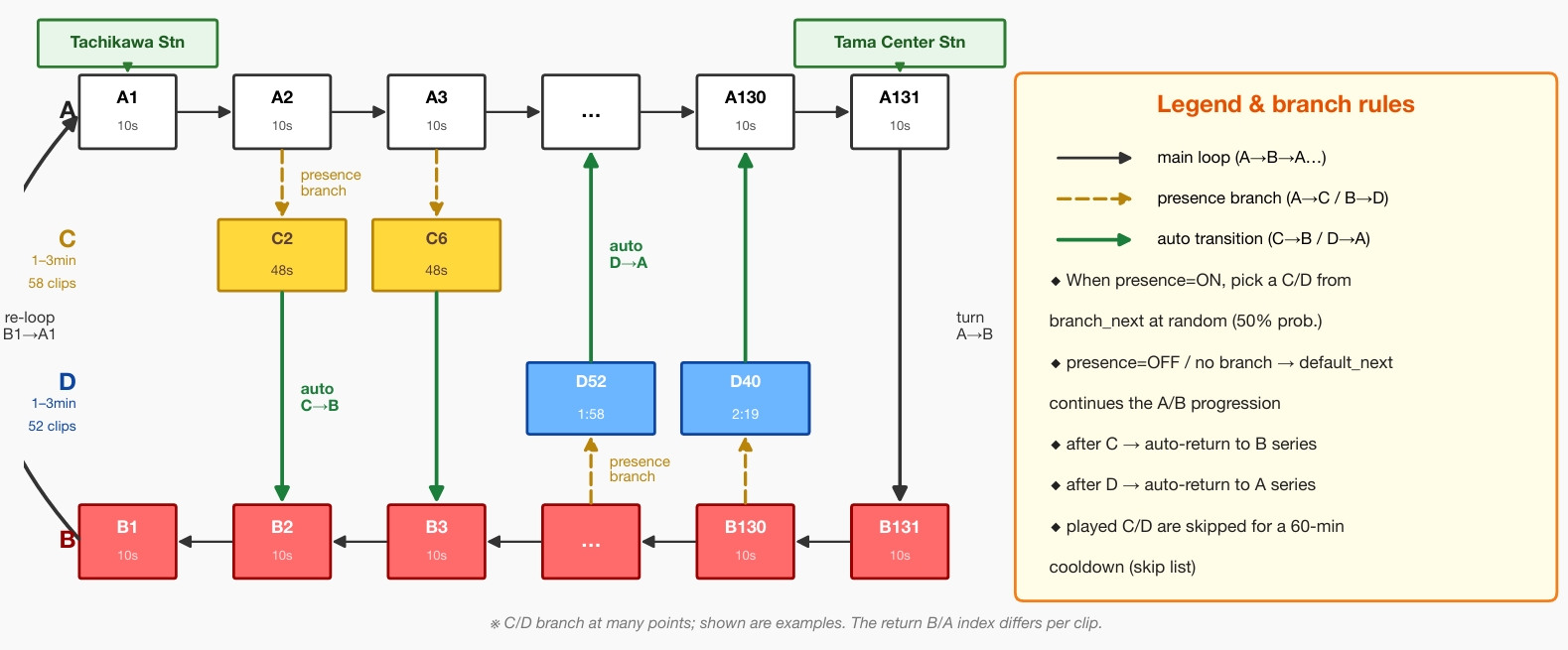}
  \caption{
  Transition graph of the rendered video sequences in \textit{Passing}. 
Groups A and B form the default loop, while Groups C and D provide viewer-presence-enabled transition branches between the two primary states. 
After a C or D clip is played, playback automatically returns to the B or A series, respectively, and recently played transition clips are temporarily excluded from the candidate pool to reduce immediate repetition.
  }
  \label{fig:sequence_transition_appendix}
\end{figure}

The visual component of \textit{Passing} is based on a rendered video dataset generated from a single continuous recording captured through the window of the Tama Monorail. The recorded route corresponds to the approximately 20 min trip from Tachikawa to Tama Center. The source footage was captured at 4K 480-fps HDR, while the transformed clips were pre-rendered as 4K HDR video for 120-fps playback in the installation.

The rendered clips are organized into four groups:
\begin{itemize}
    \item \textbf{Group A:} standard forward progression 
    (131 clips, approximately 10 s each).
    \item \textbf{Group B:} reverse playback combined with left--right mirroring 
    (131 clips, approximately 10 s each).
    \item \textbf{Groups C and D:} branching variations from Groups A and B, 
    consisting of clips in which temporal and spatial conditions are distorted 
    by digital coordinate transformation. Group C transitions from forward playback 
    toward reverse-playback and left--right-mirrored material; Group D is its inverse 
    (Group C: 58 clips; Group D: 52 clips).
\end{itemize}

These groups define both the default playback path and the possible transition branches used during installation. Fig.~\ref{fig:sequence_transition_appendix} shows the sequence-transition structure, including the default A/B loop, presence-enabled C/D branches, automatic returns after transition clips, and the temporary exclusion of recently played C/D clips.

The default loop proceeds from Group A to Group B and back to Group A: A1 through A131 are played in order, followed by B131 through B1, after which playback returns to A1. This structure allows the landscape to continue flowing from right to left while the internal temporal organization alternates between forward and reversed conditions.

Viewer presence probabilistically enables the C/D transition branches. When a viewer is detected in the viewing zone, the system may select a C or D clip from the available branch candidates instead of continuing along the default A/B progression. When no branch is selected, or when no viewer presence is detected, playback continues along the default path. After a C clip is played, playback automatically returns to the B-series; after a D clip is played, it automatically returns to the A-series. To avoid immediate repetition, C/D clips that have already been played are temporarily excluded from the candidate pool for approximately 60 min.

By combining the default A/B progression with presence-enabled C/D branches, the system produces an endless video loop from a finite source recording. The same viewer staying for an extended period is unlikely to encounter the same transition repeatedly, and each viewing session emerges as a different path through the pre-rendered spatiotemporal transformations.

\end{document}